# On direct inverse of Stokes, Helmholtz and Laplacian operators in view of time-stepper-based Newton and Arnoldi solvers in incompressible CFD

## H. Vitoshkin and A. Yu. Gelfgat

[1]School of Mechanical Engineering, Faculty of Engineering, Tel-Aviv University, Ramat Aviv 69978, Tel-Aviv, Israel

**Abstract**

Factorization of the incompressible Stokes operator linking pressure and velocity is revisited. The main purpose is to use the inverse of the Stokes operator with a large time step as a preconditioner for Newton and Arnoldi iterations applied to computation of steady three-dimensional flows and to study of their stability. It is shown that the Stokes operator can be inversed within an acceptable computational effort. This inverse includes fast direct inverses of several Helmholtz operators and iterative inverse of the pressure matrix. It is shown, additionally, that fast direct solvers can be attractive for the inverse of the Helmholtz and Laplace operators on fine grids and at large Reynolds numbers, as well as for other problems where convergence of iterative methods slows down. Implementation of the Stokes operator inverse to time-stepping-based formulation of the Newton and Arnoldi iterations is discussed.

1. **Introduction**

Motivation of this study is two-fold. First, it is connected with fully three-dimensional time-dependent CFD modelling at very large Reynolds numbers, where all the known iterative methods slow down or fail. Here we argue that for calculation on fine grids and at large Reynolds numbers the eigenvalue decomposition based direct solver [1] becomes more efficient than iterative solvers. In particular, since computational requirements of the direct solver do not depend on the time step and Reynolds number, all the time steps are completed within the same CPU time, which is an attractive feature by itself.

Second, we are interested in application of time-stepping algorithms to steady state Newton solvers and Arnoldi eigensolvers preconditioned by an inverse Stokes operator [2] below. To become an effective preconditioner the Stokes operator must be evaluated with a large time step. The latter becomes especially difficult when three-dimensional flows are studied on fine grids making most of traditional iterative methods to disconverge. In particular, we are interested in coupled incompressible pressure-velocity solvers, which are more computationally demanding than segregated ones, but possess important advantages: more stable time integration, correct calculation of pressure at each time step, and a possibility to proceed without pressure boundary conditions. Applied as preconditioners to Newton and Arnoldi solvers the coupled methods are expected to perform well if the Stokes operator with a large time step can be efficiently inversed. Considering 2D stability problems one can apply a direct sparse solver to inverse the 2D Stokes operator [4], however this becomes too memory demanding for fine three-dimensional grids. A similar approach with the same restrictions in 3D cases was implemented in [5] for explicitly calculated Jacobian matrices. At the same time, our recent pressure-velocity coupled multigrid solver [6], which performs well at small time steps fails to converge at large steps needed for 3D stability studies [4]. Based on the above experience, in this paper we recall the well-known factorization of the Stokes operator, which we use for computation of its inverse applying fast direct methods where possible. Using the finite volume method, we arrive to an analog of the Uzawa scheme [7], in which only one matrix, called "pressure matrix" has to be inversed iteratively. We show that the latter inverse can perform faster up if preconditioned by the inverse of pressure Laplacian, which is computed by a fast direct method. As a result, we arrive to a time-stepping method, which may be too CPU-time consuming for a straight-forward time-integration, but yields the inverse of the Stokes operator with a large time step, for which we are seeking.

We start the description of our approach from the factorization of the Stokes operator and the corresponding matrix resulting from the finite volume discretization. We show that the factorization can be obtained in different ways that all lead to equivalent results. Then we discuss the connection of our factorization with Chorin's projection, Uzawa and related methods. We argue that in the general case the approximations of the gradient and divergence operators are not the same which makes our pressure matrix not necessarily symmetric and positive semi definite, contrary to what is usually assumed for the Uzawa matrix. We discuss also how spatial discretization of the Stokes operator should be done to exclude pressure boundary conditions from a coupled pressure-velocity formulation.

The largest part of the direct Stokes operator inverse reduces to inverses of the Helmholtz operators. To do these inverses independently on the time step we apply the already mentioned factorization of Lynch et al. [1]. This method is based on the eigenvalue decomposition of one-dimensional operators from which the discretized operator is restored via tensor, or Kronecker, products. During many years, this method was used for higher order spectral and pseudo-spectral methods (see, e.g., [7]-[10]), but, seemingly, is very rarely applied to lower-order finite differences, finite volumes or finite elements methods [11]-[13]. We apply the eigenvalue decomposition in all directions, or treat one of the directions by the Thomas algorithm, which retains the solution direct but reduces the overall computational work. In case of a uniform grid the eigenvalue decomposition can be replaced by fast Fourier transform, and the Thomas algorithm must be replaced by cyclic reduction, which is realized in the well-known FISHPACK package [14]. We, however, consider application of uniform grids as an exceptional case, and seek a solution applicable for any non-uniform structured grid.

As a preliminary step, we examine computational performance of the above direct solver when implemented in a pressure-velocity segregated time-integration solver. We consider a series of well-known natural convection benchmarks with the purpose of comparing its performance of the direct solvers with that of an iterative method. We have chosen the BiCGstab(2) iteration as a representative example of modern Krylov-subspace-based iteration methods. Our test calculations show that the direct methods perform better than BiCGstab(2) on fine grids and for problems with large Reynolds or Grashof numbers. Since performance of the direct methods is independent on problem governing parameters and the time step size, we argue that they can be attractive for fluid flow computational modeling at large Reynolds numbers with a high spatial resolution.

The above test calculations allow us, in particular, to estimate computational cost of inverse of the Helmholtz operators needed for the inverse of the Stokes operator. The remaining step, which is done iteratively, is the inverse of the pressure matrix. We observe that this matrix is close to the pressure Laplacian and find that the inversed pressure Laplacian can be a good preconditioner. The inverse of the pressure Laplacian is also done by the fast direct method. The preconditioned BiCGstab(2) iteration converges in 2-3 iterations for small time steps and within 6-8 iterations for large ones, which we consider as an acceptable performance.

Finally, we perform a test calculation for the time-stepping based Newton steady solver and Arnoldi eigensolver. These computations show that the present approach removes the memory restrictions of the technique proposed in [5], however remains too slow on a scalar computer. In contrast to computations of [5] the present approach is scalable and can be speeded up using massively parallel calculations.

## 2. Symbolic factorization of the Stokes operator

Consider numerical semi-implicit time integration of the incompressible Navier-Stokes equations where linear pressure and velocity terms are treated implicitly and all the other terms – explicitly. Independently on a spatial discretization this leads to a system of linear algebraic equations with the Stokes operator that links velocity and pressure. For simplicity we start from a consideration of the two-dimensional case, where $u$ and $v$ are $x$- and $y$- velocity components and $p$ is the pressure. The system of equations discretized in time, but with remaining spatial differential operators can be written as

$$\begin{bmatrix} H_u & 0 & -\nabla_p^x \\ 0 & H_v & -\nabla_p^y \\ \nabla_u^x & \nabla_v^y & 0 \end{bmatrix} \begin{bmatrix} u \\ v \\ p \end{bmatrix} = \begin{bmatrix} R_u \\ R_v \\ 0 \end{bmatrix} \qquad (1)$$

Here $\nabla^x$ and $\nabla^y$ are the first derivatives in the $x$- and $y$- directions and $H = \Delta - I/\delta t$ are Helmholtz operators. $\Delta$ is the Laplacian operator, $I$ is the identity operator and $\delta t$ is the time step. The lower indices show on which variable an operator acts. The right hand sides contain the non-linear terms and all other terms that are treated explicitly. The left hand side 3×3 operator matrix assembles the 2D Stokes operator.

For any spatial discretization, we can associate the vector of unknowns in Eq. (1) with a vector assembled from all scalar unknowns of the problem, and the operators of the left hand sides as matrices containing a discretization of the corresponding operator. By assigning the lower indices, we also take into account a possibility of different discretization of different terms, which takes place, e.g., on staggered grids. Therefore, in all further considerations $H_u \neq H_v$, $\nabla_u^x \neq \nabla_p^x$, and $\nabla_u^y \neq \nabla_p^y$.

Treating the Stokes operator as a 3×3 matrix, we derive its $LU$ decomposition. Assigning identity operators to the main diagonal of either $L$ or $U$ we arrive to the following expressions:

$$\begin{bmatrix} H_u & 0 & -\nabla_p^x \\ 0 & H_v & -\nabla_p^y \\ \nabla_u^x & \nabla_v^y & 0 \end{bmatrix} = \begin{bmatrix} I & 0 & 0 \\ 0 & I & 0 \\ \nabla_x^u H_u^{-1} & \nabla_y^u H_v^{-1} & I \end{bmatrix} \begin{bmatrix} H_u & 0 & -\nabla_p^x \\ 0 & H_v & -\nabla_p^y \\ 0 & 0 & C \end{bmatrix} \qquad (2)$$

$$\begin{bmatrix} H_u & 0 & -\nabla_p^x \\ 0 & H_v & -\nabla_p^y \\ \nabla_u^x & \nabla_v^y & 0 \end{bmatrix} = \begin{bmatrix} H_u & 0 & 0 \\ 0 & H_v & 0 \\ \nabla_u^x & \nabla_v^y & C \end{bmatrix} \begin{bmatrix} I & 0 & -H_u^{-1}\nabla_p^x \\ 0 & I & -H_v^{-1}\nabla_p^y \\ 0 & 0 & I \end{bmatrix} \qquad (3)$$

where

$$C = \nabla_u^x H_u^{-1} \nabla_p^x + \nabla_v^y H_v^{-1} \nabla_p^y \tag{4}$$

Another possibility is to apply the Sherman-Morrison-Woodbury (SMW) formula. Denoting the Stokes operator as $S$ we define

$$S = G + VW^T, \quad G = \begin{bmatrix} H_u & 0 & 0 \\ 0 & H_v & 0 \\ \nabla_u^x & \nabla_v^y & I \end{bmatrix}, \quad V = \begin{bmatrix} \nabla_p^x \\ \nabla_p^y \\ I \end{bmatrix}, \quad W = \begin{bmatrix} 0 \\ 0 \\ -I \end{bmatrix}, \tag{5}$$

and according to the SMW formula

$$S^{-1} = (G + VW^T)^{-1} = G^{-1} - G^{-1}VCW^TG^{-1} = G^{-1}(I - VCW^TG^{-1}) \tag{6}$$

Simple evaluations yield

$$G^{-1} = \begin{bmatrix} H_u^{-1} & 0 & 0 \\ 0 & H_v^{-1} & 0 \\ -H_u^{-1}\nabla_u^x & -H_v^{-1}\nabla_v^y & I \end{bmatrix} \tag{7}$$

and the matrix $C$ is defined by Eq. (4). Other factorization methods, e.g., Schur complement equation [15]-[17], lead to an equivalent formulation. The matrix $C$ is an analog of the Schur complement matrix (or Uzawa matrix), which arises in the well-known Uzawa method [7],[17]. Our matrix $C$, however, is different since we took into account that velocity divergence and pressure gradient operators on staggered grids act on variables defined at different nodes and also result at different nodes. Therefore, these operators cannot be connected via the transpose operation. Additionally, computation of the divergence in the whole flow region involves velocity boundary values, while computation of the pressure gradient does not use pressure boundary values. Contrarily to the Uzawa matrix, which is symmetric and positive semi defined [17], the matrix $C$ is not necessarily symmetric and its positive definition should be examined for each scheme separately.

It is well known that the Uzawa method usually is not applied directly, but is used as a starting point for definition of various segregated pressure-velocity solvers [7],[17]. We, however, are interested in the implementation of the Stokes operator factorization directly. As mentioned, the equations (2) – (7) define three equivalent ways to derive the solution of Eqs. (1). The solution is obtained in three steps:

1. Solve $\hat{u} = H_u^{-1} R_u$ and $\hat{v} = H_v^{-1} R_v$ for $\hat{u}$ and $\hat{v}$.

2. Solve $p = -C^{-1}\left(\nabla_x^u \hat{u} + \nabla_y^v \hat{v}\right)$ for $p$.

3. Solve $u = \hat{u} + H_u^{-1}\nabla_p^x p$ and $v = \hat{v} + H_v^{-1}\nabla_p^y p$.

Thus, calculation of the solution of a 2D problem (1) requires four inverses of the Helmholtz operator and one inverse of the matrix $C$. Since the matrix $C$ defines solution for the pressure, we call it "pressure matrix". Extension to a three-dimensional case is straight-forward. The system of equations is defined as

$$\begin{bmatrix} H_u & 0 & 0 & -\nabla_p^x \\ 0 & H_v & 0 & -\nabla_p^y \\ 0 & 0 & H_w & -\nabla_p^z \\ \nabla_u^x & \nabla_v^y & \nabla_w^z & 0 \end{bmatrix}\begin{bmatrix} u \\ v \\ w \\ p \end{bmatrix} = \begin{bmatrix} R_u \\ R_v \\ R_w \\ 0 \end{bmatrix} \tag{8}$$

and the *LU* decomposition of the Stokes operator becomes, e.g.,

$$\begin{bmatrix} H_u & 0 & 0 & -\nabla_p^x \\ 0 & H_v & 0 & -\nabla_p^y \\ 0 & 0 & H_w & -\nabla_p^z \\ \nabla_u^x & \nabla_v^y & \nabla_w^z & 0 \end{bmatrix} = \begin{bmatrix} I & 0 & 0 & 0 \\ 0 & I & 0 & 0 \\ 0 & 0 & I & 0 \\ \nabla_u^x H_u^{-1} & \nabla_v^y H_v^{-1} & \nabla_w^z H_w^{-1} & I \end{bmatrix}\begin{bmatrix} H_u & 0 & 0 & -\nabla_p^x \\ 0 & H_v & 0 & -\nabla_p^y \\ 0 & 0 & H_w & -\nabla_p^z \\ 0 & 0 & 0 & C \end{bmatrix}, \tag{8}$$

$$C = \nabla_u^x H_u^{-1}\nabla_p^x + \nabla_v^y H_v^{-1}\nabla_p^y + \nabla_w^z H_w^{-1}\nabla_p^z. \tag{9}$$

As in the 2D case the solution is calculated in the similar three steps

1. Solve $\hat{u} = H_u^{-1}R_u$, $\hat{v} = H_v^{-1}R_v$ and $\hat{w} = H_w^{-1}R_v$ for $\hat{u}, \hat{v}$ and $\hat{w}$.

2. Solve $p = -C^{-1}\left(\nabla_x^u \hat{u} + \nabla_y^v \hat{v} + \nabla_z^v \hat{w}\right)$ for $p$.

3. Solve $u = \hat{u} + H_u^{-1}\nabla_p^x p$, $v = \hat{v} + H_v^{-1}\nabla_p^y p$, and $w = \hat{w} + H_w^{-1}\nabla_p^z p$.

Calculation of a 3D solution requires 6 inverses of the Helmholtz operator and one inverse of the pressure matrix $C$. Note, that if to assume very small time step $\delta t$, and at the steps 2 and 3, as well as in Eqs. (4) and (9) apply $H \approx I/\delta t$, then operator $C$ turns into the Laplacian of pressure, and step 3 can be interpreted as a projection of intermediate solution $(\hat{u}, \hat{v}, \hat{w})$ on the divergence-free space. Thus, we arrive to the standard Chorin's projection method.

For the fully coupled implementation of the steps 1-3, one needs inverse of the Helmholtz operators and of the pressure matrix $C$. The inverse of the Helmholtz operators is usually a part of a pressure-velocity segregated code. Since at small time steps, the Helmholtz operator is close to the identity operator, its iterative inverse typically does not involve any numerical difficulties.

As it was mentioned above, in some applications involving Newton and Arnoldi iterations for computation of steady solutions and analysis of their stability, the inverse Stokes operator with a large time step is used as a preconditioner [1],[3]. In these cases an iterative inverse of the Helmholtz operators can be problematic. In such cases, the direct methods discussed in the next Section can be called for.

The main difficulty in the implementation of the proposed Stokes operator inverse is computation of the inversed pressure matrix *C*. It is easy to see that this matrix is singular, which is a usual consequence of the pressure defined up to an additive constant. As we already mentioned, at small time steps this matrix is close to approximation of the pressure Laplacian, thus causing well known problems when it is inverted. Apparently, a Dirichlet point should be added to make the matrix regular. The pressure boundary conditions, however, can be avoided if calculation of the pressure gradient in step 3 does not involve pressure boundary values. The simplest example of that is the use of staggered grids, as was proposed by Patankar and Spalding [18], and was implemented, e.g., in our earlier studies [5],[6]. In general, to make pressure boundary conditions unnecessary the low-order numerical scheme must: (i) close the system of the momentum equation by the continuity equation ($div\boldsymbol{v} = 0$ in the non-boundary points) and not by a pressure or pressure correction equation that always requires boundary conditions; and (ii) approximation of the pressure gradient in the momentum equations should not involve pressure boundary values.

Since computation of the whole matrix *C* can be CPU-time and memory consuming, the most natural way of its inverse is implementation of one of the Krylov subspace iteration methods, which requires only calculation of multiplication of the matrix by a vector (the action of a matrix). The latter can be done by using Eqs. (4) or (9) at the cost of two and three Helmholtz operator inverses for 2D and 3D problems, respectively. Our numerical experiments, (see below for a detailed description) showed that inverse of *C* on the $100^2$ stretched grid by the BICGstab(2) method requires 80-100 iterations. At small time steps, the Helmholtz operators in Eqs. (4) and (9) tend to the identity operator, so that the whole matrix *C* tends to the approximation of the Laplacian of pressure. Thus, at small time steps the number of iterations can be significantly decreased by use of the inverse pressure Laplacian matrix as a preconditioner. The latter makes the resulting matrix close to the unity, so that the BICGstab(2) method converges in less than 2-4 iterations for a dimensionless time step of 0.01, and in 6-8 iterations if the time step is increased to the value of 1. The verification of the above three step algorithms against some known benchmark results is straight-forward and is not reported here. It is emphasized that evaluation of the action of matrix *C* requires two or three inverses of the Helmholtz operators in 2D and 3D cases, respectively. Therefore it is a CPU-time consuming operation, so that time integration is affordable only if *C* can be inversed in very few iterations.

It is clear at this stage that implementation of the above algorithm to the Stokes operator inverse for time-dependent computations will be much more computationally demanding than most pressure-velocity segregated techniques. A comparison with the latter, if meaningful at all, is beyond the scope of the present paper. It should be emphasized however, that if direct methods are applied to the inverse of the Helmholtz operators (see below), then the inverse of the pressure matrix *C* remains the only iterative part of the algorithm. We call it semi-direct inverse of the Stokes operator.

A weak dependence of the proposed Stokes operator inverse on the time step allows one to perform calculations with large time steps, which we consider as a prerequisite to applications of methodology of [1],[3] to calculate developed steady three-dimensional flows and to study their stability. In the following, we describe this approach using the present notations. Assuming $U$ as a vector containing all the unknown values of pressure and velocity, a steady solution of the discretized Navier-Stokes equation is defined by

$$F(U) = LU + N(U) + f = 0 \qquad (10)$$

where $L$ stays for the linear operators, $N$ for the non-linear terms and $f$ for all the additional volume forces. The Newton iteration for equations (10) is defined as

$$F_U(U^{(n)})u = F(U^{(n)}) \qquad (11)$$

$$U^{(n+1)} = U^{(n)} - u \qquad (12)$$

Here $F_U$ is the Jacobian matrix of (10) and $u$ is the correction calculated as a solution of the linear equation system (11). Apparently the stable steady state of (10) can be calculated via a time-integration process

$$\frac{U^{k+1}-U^k}{\delta t} = LU^{k+1} + N(U^k) + f \qquad (13)$$

Taking into account that $L - I/\delta t = -S$, where S is the Stokes operator defined in Eq. (1), Eq. (13), this can be rewritten as

$$U^{k+1} - U^k = -S^{-1}F(U^k) \qquad (14)$$

Similarly, considering the linearized time step

$$\frac{u^{k+1}-u^k}{\delta t} = Lu^{k+1} + N_U u^k \qquad (15)$$

we arrive to the relation

$$u^{k+1} - u^k = -S^{-1} F_U u^k \qquad (16)$$

Rewriting Eq. (11) in an equivalent form

$$-S^{-1} F_U(U^{(n)}) u = -S^{-1} F(U^{(n)}) \qquad (17)$$

We observe that the r.h.s of Eq. (17) can be calculated as a difference between two consecutive time steps (14), while action of the l.h.s. matrix $-S^{-1} F_U(U^{(n)})$ on an arbitrary vector $u^k$ is defined as a difference between two consecutive linearized time steps (16). The latter allows one to apply the Krylov-subspace-based iteration methods to calculate the Newton correction $u$.

For a steady state $U$, the linear stability problem is reduced to the generalized eigenvalue problem

$$F_U(U)v = [L + N_U(U)]v = \lambda B v, \qquad (18)$$

where $\lambda$ is the eigenvalue, $v$ is the eigenvector, and $B$ is the diagonal matrix whose diagonal values are unity for the rows corresponding to the velocities and zeroes for those corresponding to the pressure. Since the matrix $B$ is singular, and the Arnoldi iterations converge effectively to the eigenvalues with the largest absolute value [5], it is necessary to consider the eigenproblem in the shift-and inverse transformation mode

$$\mu v = [L + N_U(U) - \sigma I]^{-1} v, \quad \mu = \frac{1}{\lambda - \sigma} \qquad (19)$$

where $\sigma$ is a complex shift which must be chosen close to the leading eigenvalue $\lambda$. Application of the Arnoldi iteration requires computation of Krylov basis vectors defined as

$$v^{n+1} = [L + N_U(U) - \sigma I]^{-1} v^n \quad \text{or} \quad [L + N_U(U) - \sigma I] v^{n+1} = v^n \qquad (20)$$

It is easy to see that

$$-S^{-1} v^{n+1} = -S^{-1}[L + N_U(U) - \sigma I] v^n = v^n(t + \Delta t) - v^n(t) + \sigma S^{-1} v^n(t) \qquad (21)$$

The term $S^{-1} v^n(t)$ can be interpreted as a Stokes time step with the nullified forcing, i.e., with $N=f=0$. Thus, instead of the second equality of (20), we consider the equivalent one

$$-S^{-1}[L + N_U(U) - \sigma I] v^{n+1} = -S^{-1} v^n \qquad (22)$$

The r.h.s. of Eq. (22) is calculated as the Stokes time step with the nullified forcing. The action of the l.h.s. matrix on a vector is calculated according to Eq. (21) as the difference between two consecutive linearized time steps to which the linearized Stokes time step multiplied by a shift should be added. This allows us to apply a Krylov subspace iteration method for calculation of the next Krylov vector $\boldsymbol{v}^{n+1}$ for the Arnoldi iteration.

Note that the inverse Stokes operator $\boldsymbol{S}^{-1}$ serves as a preconditioner in Eqs. (17) and (22). At small time steps the Helmholtz operators in Eq. (1) tend to unity, so that $\boldsymbol{S}^{-1}$ will affect mainly the pressure terms, from which we cannot expect any convergence improvement. However, application of $\boldsymbol{S}^{-1}$ with a large time step as a preconditioner will make the product $\boldsymbol{S}^{-1}\boldsymbol{L}$ close to the unity matrix that may improve the convergence.

### 3. Tensor-product inverse of Laplacian and Helmholtz operators revisited

In this Section we revisit the result of Lynch et al. [1] with the aim to show that with the mesh refinement, and especially for 3D computations, the tensor-product based direct inverse of the matrices resulting from Laplacian and Helmholtz operators can become faster than traditional iterative techniques.

Consider a Laplace operator acting on a scalar function $u(x,y)$, defined on a rectangle $0 \leq x \leq a, 0 \leq y \leq b$. The function $u$ satisfies Dirichlet, Neumann or mixed linear homogeneous boundary conditions. We assume that the rectangle $0 \leq x \leq a, 0 \leq y \leq b$ is covered by an orthogonal grid, whose nodes are defined as $x_i$ and $y_j$, $i = \overline{1, N_x}, j = \overline{1, N_y}$. We denote discretization of the $x$- and $y$- second derivatives as operators $D_{xx}$ and $D_{yy}$, that are one-dimensional and act on a row or a column of the grid function $u_{ij}=u(x_i,y_j)$, respectively. Representing $D_{xx}$ and $D_{yy}$ by matrices and following notations of the Kronecker product we write the discretized Poisson equation as

$$\Delta u = [D_{xx} \otimes I_y + I_x \otimes D_{yy}]u = f \tag{23}$$

where $I_x$ and $I_y$ are identity matrices of the order $N_x$ and $N_y$, respectively, $\otimes$ denotes the tensor product, and $f_{ij}=f(x_i,y_j)$ is the discretized right hand side of the Poisson equation. For the following we assume that the eigenvalue decompositions of matrices $D_{xx}$ and $D_{yy}$ are known and are represented as

$$D_{xx} = E_x \Lambda_x E_x^{-1}, \quad D_{yy} = E_y \Lambda_y E_y^{-1}. \tag{24}$$

Here $E_x$ and $E_y$ are square matrices of the order $N_x$ and $N_y$, respectively, whose columns are eigenvectors of the matrices $D_{xx}$ and $D_{yy}$. $\Lambda_x$ and $\Lambda_y$ are diagonal matrices having the eigenvalues of $D_{xx}$ and $D_{yy}$ on their diagonals. According to [1] the solution of Eq. (23) can be represented as

$$u = (E_x \otimes E_y)\Lambda^{-1}(E_x^{-1} \otimes E_y^{-1})f \qquad (25)$$

where

$$\Lambda = (\Lambda_x \otimes I_y) + (I_x \otimes \Lambda_y) \qquad (26)$$

is a diagonal matrix of the order $N_x N_y$ whose diagonal values are $\Lambda_{ij} = \Lambda_{x,i} + \Lambda_{y,j}$.

In the case of the Helmholtz equation

$$(\Delta + aI)u = [D_{xx} \otimes I_y + I_x \otimes D_{yy} + aI_x \otimes I_y]u = f \qquad (27)$$

the solution $u$ can be expressed as in Eq. (25) with $\Lambda_{ij} = \Lambda_{x,i} + \Lambda_{y,j} + a$.

Finally, in the case of the Neumann problem for the Poisson equation we define

$$\Lambda_{ij} = \begin{cases} \Lambda_{x,i} + \Lambda_{y,j} & if\ \Lambda_{x,i} + \Lambda_{y,j} \neq 0 \\ 1, & otherwise \end{cases} \qquad (28)$$

It can be shown that a replacement of the zero value of $\Lambda_{ij}$ by unity is equivalent to definition of a zero Dirichlet point at the boundary.

The eigenvalue decompositions (24) of the one-dimensional operators $D_{xx}$ and $D_{yy}$ are computed in $O(N_x^3)$ and $O(N_y^3)$ operations, respectively. For a time–marching procedure (see below) this computation is needed only once, so that its computational cost can be neglected if the number of time steps is sufficiently large. Once these are known, the calculation of the solution of Eq. (10) or (14) requires $2N_x N_y(N_x+N_y)$ multiplications and $N_x N_y$ divisions by $\Lambda_{ij}$ for calculations of $u$ from Eq. (12). Therefore the total amount of multiplications and divisions is $N_x N_y(2N_x+2N_y + 1)$. This can be compared with the computational cost of the Gauss elimination of a banded matrix, which for the second-order finite difference discretization of the Laplacian has band width $M = 2N_x$. Then the preprocessing step, say the $LU$-decomposition, will need $O\left(M(N_x N_y)^2\right)$ operations, and $O(M^2 N_x N_y)$ operations for each individual solution, which is much larger than that for the proposed method. In the special case of uniform grid the Fourier transform can replace the eigenvalue decomposition. Then multiplication by the operators $E_x$ or $E_y$ requires $N_x log_2 N_x$ and $N_y log_2 N_y$ operations, reducing the operation count to $(N_x N_y)(log_2 N_x + log_2 N_y)$ [14]. Nevertheless, iterative methods may still converge in a lesser amount of operation, so that the choice of a solution method should always be checked by representative test calculations.

The generalization of (25) for three-dimensional case is straight-forward. Adding the third eigenvalue decomposition $D_{zz} = E_z \Lambda_z E_z^{-1}$, we obtain

$$u = (E_x \otimes E_y \otimes E_z) \Lambda^{-1} (E_x^{-1} \otimes E_y^{-1} \otimes E_z^{-1}) f \tag{29}$$

where $\Lambda_{ijk} = \Lambda_{x,i} + \Lambda_{y,j} + \Lambda_{z,k}$. The whole procedure needs $N_x N_y N_z (3N_x + 3N_y + 3N_z + 1)$ multiplications and divisions. In the following we refer to Eqs. (25) and (29) as the tensor product factorization (TPF) solver proposed in [1]. As mentioned, this solver is often applied together with spectral and pseudospectral methods [8]-[10], however its application together with lower-order spatial discretization, remains rare [11]-[13]. Regarding the lower-order methods, two additional comments should be made. First, increase of an approximation order by use of longer stencils will not increase the computational cost of implementation of Eqs. (25) or (29), however performance of any iterative methods will be affected due to lesser sparseness of the matrices. Second, short three- or five-points stencils usually used in lower order methods allow one to replace the eigenvalue decomposition in one of directions by the Thomas algorithm, thus applying the Haidvogel-Zang decomposition [19]. This will retain the direct inverse of the matrices, but will decrease the overall computational time. In the test computations below, we call this approach the TPT solver. Apparently, the direction with a maximal number of grid points should be chosen for such a replacement. The Thomas algorithm applied for *N* grid points and a scheme defined on the 3-point stencil requires 5*N* multiplications and divisions, which is significantly less than $N^2$ multiplications needed for computations of the mass-vector product. Assuming in the above $N_x=N_y=N_z>>5$ we see that application of the Thomas algorithm in one direction reduces the number of operations almost twice for a 2D case and by a factor of approximately 2/3 for a 3D case.

4. **Computation of the pressure matrix using the tensor-products**

To obtain the analytical expression for the pressure matrix *C,* we define one-dimensional operators *D* describing the first derivative operators of Eq. (4):

$$\nabla_u^x = D_u^x \otimes I_y \otimes I_z, \quad \nabla_v^y = I_x \otimes D_v^y \otimes I_z \tag{30}$$

$$\nabla_p^x = D_p^x \otimes I_y \otimes I_z, \quad \nabla_p^y = I_x \otimes D_p^y \otimes I_z \tag{31}$$

Then, using Eq. (25) for the inverse operators and performing some simple evaluations with the Kronecker products we obtain the following expressions for the two terms of Eq. (4):

$$\nabla_x^u H_u^{-1} \nabla_x^p = D_x^u \left[ (E_x^u \otimes E_y^u) \left( -\tau^{-1} I_x^u \otimes I_y^u + \Lambda_x^u \otimes I_y^u + I_x^u \otimes \Lambda_y^u \right)^{-1} \left( E_x^{u-1} D_x^p \otimes E_y^{u-1} \right) \right] \quad (32)$$

$$\nabla_y^v H_v^{-1} \nabla_y^p = D_y^v \left[ (E_y^v \otimes E_x^v)^T \left( -\tau^{-1} I_x^v \otimes I_y^v + \Lambda_x^v \otimes I_y^v + I_x^v \otimes \Lambda_y^v \right)^{-1} \left( E_x^{v-1} \otimes E_y^{v-1} D_y^p \right) \right] \quad (33)$$

and derivation of the similar expressions for the three-dimensional case is straight-forward.

The expressions (4), (32), and (33) form the analytical representation of the pressure matrix *C*. Formally, it can be calculated and inversed, which completes a fully direct inverse of the Stokes operator. Practically, evaluation of the above Kronecker products is CPU-time consuming, but possible. The matrix *C* is not sparse, so that its direct inverse is practically impossible. Relations (32) and (33) can be used also for evaluation of the action of matrix *C*.

Calculating the matrix *C* and observing its components we find that many of them are, in fact, numerical (but not analytical) zeroes. After nullifying all the components that are at least 10 orders of magnitude smaller than the leading term of the same row, we discover that the matrix *C* can be approximated by a sparse matrix. This artificial sparseness is most profound when the uniform grid is used, for which more than 95% of the matrix elements are numerical zeroes. This allows us to apply a sparse matrix solver and to keep the *LU* decomposition of the resulting matrix. Then carrying out of step 2 of the above algorithms reduces to computation of sparse back/forward substitutions, which makes the inverse of the Stokes operator fully analytical. For stretched grids, this approach is possible, however not efficient. Some additional details are given below.

**5. Implementation of TPF and TPT for incompressible CFD time marching**

For test calculations we consider a system of Boussinesq equations describing convection in a rectangular 3D box. The flow is described by the energy, momentum and continuity equations (details can be found in [16]-[26])

$$\frac{\partial T}{\partial t} + (\boldsymbol{v} \cdot \nabla) T = \frac{1}{PrGr^{1/2}} \Delta T \quad (34)$$

$$\frac{\partial \boldsymbol{v}}{\partial t} + (\boldsymbol{v} \cdot \nabla) \boldsymbol{v} = -\nabla p + \frac{1}{Gr^{1/2}} \Delta \boldsymbol{v} + T \boldsymbol{e}_z \quad (35)$$

$$div\, \boldsymbol{v} = 0 \quad (36)$$

where $\boldsymbol{v} =(u,v,w)$, $p$, $t$, and $T$ are the dimensionless velocity, pressure, time and temperature, respectively, and $\boldsymbol{e}_z$ is the unit vector in $z$-direction. It follows from Eq. (18) that in the formulation used $Gr^{1/2}$ yields an estimation of the Reynolds number.

As an example of numerical solution of eqs. (34)-(36) a standard incremental pressure-correction scheme [[20]] is applied. Denoting the time step by $\delta t$ and by the superscript $(n)$ the values of the functions at $t=n\delta t$, we perform the time integration as

$$\frac{1}{2\delta t}\left(3T^{(n+1)} - 4T^{(n)} + T^{(n-1)}\right) + \left(\boldsymbol{v}^{(n)} \cdot \nabla\right)T^{(n)} = \frac{1}{PrGr^{1/2}}\Delta T^{(n+1)} \qquad (37)$$

$$\frac{1}{2\delta t}\left(3\boldsymbol{v}^{(n+1/2)} - 4\boldsymbol{v}^{(n)} + \boldsymbol{v}^{(n-1)}\right) + \left(\boldsymbol{v}^{(n)} \cdot \nabla\right)\boldsymbol{v}^{(n)} = -\nabla p^{(n)} + \frac{1}{Gr^{1/2}}\Delta \boldsymbol{v}^{(n+1/2)} + T^{(n+1)}\boldsymbol{e}_z \qquad (38)$$

$$\Delta(\delta p) = -\frac{1}{\delta t} div\, \boldsymbol{v}^{(n+1/2)} \qquad (39)$$

$$p^{(n+1)} = p^{(n)} + \delta p, \quad \boldsymbol{v}^{(n+1)} = \boldsymbol{v}^{(n+1/2)} + (\delta t) grad(\delta p) \qquad (40)$$

Extracting the problems for $T^{(n+1)}$ from Eq. (37) and $\boldsymbol{v}^{(n+1/2)}$ from Eq. (38), performing one time step reduces to the solution of one Helmholtz equation for the temperature, three Helmholtz equations for the velocity components and one Poisson equation for the pressure, in addition to the calculation of the nonlinear advective terms. Note that also more advanced projection schemes or SIMPLE-like algorithms (see, e.g., [20],[21]) also consist primarily of a sequence of Helmholtz and Poisson problems, so that the following conclusions apply also in all these cases.

In the following, to illustrate when the eigenvalue decomposition approach may be more effective than an iterative solution, we consider several benchmark problems on natural convection in laterally heated two-and three-dimensional cavities [9],[22],[24]-[26]. We solve the Helmholtz and Poisson equations both by Jacobi (diagonal) preconditioned BiCGstab(2) iteration [22], by the tensor product factorization (TPF) method, and by the tensor product method combined with the Thomas algorithm (TPT) in one of the directions. In all the results reported both approaches yielded numerical solutions that coincided at least to within the tenth decimal digit. After establishing the equivalence of all the three solutions we compare the consumed CPU times.

The choice of Jacobi preconditioner for the BiCGstab(2) iteration is justified by diagonal dominance of the matrices and its negligible computational cost (see also [23]). We are aware of the fact that Krylov subspace iteration methods, like BiCGstab(2), with a smarter choice of a preconditioner, can perform faster than they do in the following test calculations. However, the choice of a preconditioner is usually problem-dependent, the effect we want to avoid. Moreover, we believe that the qualitative conclusions we derive will hold also in the case of more efficient iteration techniques applied to large Reynolds number CFD problems.

## 5. Test calculations

### 5.1. TPF and TPT solvers versus BiCGstab(2)

As mentioned above, to compare CPU time consumption of the fast direct solvers, $t_{TPF}$ and $t_{TPT}$, with that of the BiCGstab(2) iteration, $t_{BiCG}$, we consider well-known benchmark problems on natural convection in laterally heated rectangular and three-dimensional cavities. In the following, we perform the time integration based on Eqs. (34)-(36) with the finite volume discretization in space. We consider convection of air ($Pr$=0.71) in a laterally heated square cavity [24],[25], in a two-dimensional cavity with height-to-width ratio $A$=8 [26], and in a laterally heated cubical box [9]. The finite volume staggered grid is stretched near the boundaries.

It is emphasized that the numerical method and the code are already completely verified [5],[6]. Here we are interested only in comparison of consumed CPU times. To do that we start from the laterally heated square cavity and perform time-dependent calculations for $Gr=10^5$ until convergence to a steady state. Then we set the Grashof number to $Gr=10^6$, use the calculated steady state as an initial condition, and carry out 10,000 time steps. Then we again increase the Grashof number by an order of magnitude to $Gr=10^7$, and perform 10,000 more time steps. All the runs were carried out on a $100^2$ nodes grid with the time step $\delta t$=0.01, using either the BiCGstab(2), TPF or TPT method. The consumed CPU times are reported in Table 1. To gain additional information we also considered the supercritical oscillatory flow in a tall vertical cavity [26] and performed the calculation over 5 oscillation periods. The calculation is carried out on the grid with 100×800 nodes, for $A$=8 and $Gr=4.8\times10^7$. To perform calculations on an already converged limit cycle solution, we used a snapshot of the oscillatory flow computed in [6] on the same grid as the initial state. The time step was $\delta t$=0.001.

Table 1 shows that the TPF and TPT methods consume approximately the same CPU time for all of the unknown functions: $u$, $v$, $T$ (which satisfy a Dirichlet problem for the Helmholtz equation) and $\delta p$ (which satisfies a Neumann problem for the Poisson equation). The TPT method is faster in agreement with the above estimations. The slight difference in $t_{TPF}$ and $t_{TPT}$ times consumed for different variables is an effect of the staggered grid that has a slightly different number of nodes for each of the unknowns. In contrast, BiCGstab(2) converges much faster for the Helmholtz problems than for the Poisson problem. This is because the Helmholtz operator, which is close to being a perturbation of the identity operator, is far better conditioned than the Laplacian, and hence requires many fewer BiCGstab(2) iterations to converge. More specifically, BiCGstab(2) requires between 10 and 33 times longer to solve the Poisson problem than to solve one of the Helmholtz problems. In consequence, BiCGstab(2) is about 3 to 5 times *faster* than TPF for each of the Helmholtz problems and about 5 times *slower* than TPF for the Poisson problem. Moreover, with the increase of the Grashof number, the BiCGstab iterations converge slower, which is quite expected, while the CPU time consumptions of the TPF method does not change. Therefore, one can expect that for the flows with larger Grashof (or Reynolds) numbers, the TPF (or TPT) approach can become even more attractive. The results obtained for 2D problems (Table 1) suggest combining BiCGstab (or another iterative solver) to calculate the temperature and velocity with the TPF or TPT solver to calculate the pressure.

In the test calculations for the three-dimensional problem, we also started from $Gr=10^5$. After carrying out 10,000 time steps, with the time step $\delta t=0.001$, we increased the Grashof number to $Gr=10^6$ and then, after another 10,000 time steps to $Gr=10^7$. These calculations were performed for stretched grids consisting of $50^3$, $75^3$ and $100^3$ nodes. To explore the potential scalability of the TPF and TPT approaches we carried out these computations twice, using either scalar or vector processors. The results are summarized in Tables 2 and 3. In the three-dimensional calculations, the TPF approach is always faster than BiCGstab. For a scalar processor, the ratio $t_{BiCG}/t_{TPF}$ of CPU times is between 1.5 (for a Helmholtz problem) and 12 (for a Poisson problem), while for a vector processor, this ratio is between 2.5 and 50. As expected, the CPU time consumed by BiCGstab(2) increases with $Gr$, as well as with the grid refinement. The CPU time consumed by the TPF and TPT approaches is Grashof-number (or Reynolds-number) independent and grows with the mesh refinement according to the operation counts discussed in Section 3. We observe also that the ratios $t_{BiCG}/t_{TPF}$ and $t_{BiCG}/t_{TPT}$ obtained on a vector processor are significantly larger than the ratio obtained for a scalar processor. The proportion between the ratios corresponding to the scalar and vector processors for $50^3$ and $75^3$ grids varies between the values 3 and 4, which corresponds to the length of vector (equal to 4) in the processor used (cf. Tables 2 and 3). This ratio is larger for a finer $75^3$ grid, which also can be expected. The ratios $t_{BiCG}/t_{TPF}$ and $t_{BiCG}/t_{TPT}$ obtained on a vector processor (Table 3) for $75^3$ grid is larger than that obtained for a $100^3$ grid, so that the grid refinement does not necessarily lead to a more profound difference in the consumed CPU times. At the same time in all the 3D cases considered, the TPF and TPT methods perform significantly faster than BiCGstab(2), and the CPU time ratios $t_{BiCG}/t_{TPF}$ and $t_{BiCG}/t_{TPT}$ grow with the increase of the Grashof (Reynolds) number. This shows that time-dependent fully three-dimensional calculations possibly should use fast direct solvers instead of iterative ones.

*5.2. Implementation of the Stokes operator inverse*

The CPU time consumed when the time integration was carried out using the proposed Stokes operator inverse is reported in Table 4. The integration was repeated twice with the inverse of the pressure matrix *C* using BiCGstab(2) and by a direct inverse using the sparse matrix solver MUMPs [27]. We report the total CPU time consumed by the whole computation in each of the cases, $t_{total\_bicg\_C}$ and $t_{total\_mumps}$, as well as CPU times consumed for the inverse of the matrix *C*. The latter we denote as $t_{bicg\_C}$ for the case of the BiCGstab(2) solver. In the case of the direct MUMPs solver we report the time needed to perform the *LU* decomposition of *C*, $t_{LU}$, and the time spent for the back and forward substitution, $t_{bs}$, at the solution phase. Comparing with the CPU times reported in Table 1 we observe that, as expected, the time stepping using the proposed Stokes operator inverse is rather slow. The implementation of the direct sparse solver in this case leads to extremely slow computations and therefore is excluded from further test runs. The decrease of the CPU time consumed by MUMPs with the increase of the Reynolds number is explained by filtering out the numerical zeroes from the resulting *C* matrix, as it was explained above. Implementation of the BiCGstab(2) algorithm for the inverse of *C* shows that it requires most of the whole computational effort. As expected, the convergence of BiCGstab(2) slows down with the increase of Grashof (Reynolds) number.

Performance of the Newton method based on the eqs. (11), (12) and (17) was tested on the same 2D thermal convection benchmark [24],[25]. The steady state at $Gr=5\times10^6$ was used as an initial state to calculate steady state solution at $Gr=10^7$. The whole Newton process converges in 6 iterations. The time step $\delta t$ was varied between 1 and 100, however, for $\delta t \geq 80$ the iterations for the inverse of *C* disconverge. The total consumed CPU time, the number of BiCGstab(2) iterations needed for calculation of the current Newton correction ***u*** and the maximal number of BiCGstab(2) iterations needed for the inverse of pressure matrix *C,* are reported in Table 5. We observe that starting from $\delta t=2$ the whole Newton iteration process completes in 700-800 seconds. Clearly, the number of iterations and the consumed CPU time are problem dependent, making it impossible to find a generally optimal time step. Thus, for example, calculation of the steady state at $Gr=10^8$ using the solution at $Gr=5\times10^6$ and $\delta t=10$ as an initial guess consumes more than 2 CPU hours. The same Newton process as reported in Table 5, implemented with the approach proposed in [5], completes in less than 5 seconds on the same processor. Clearly the present approach is more than 100 times slower. At the same time, compared to the approach of [5], it has two important advantages. First, it removes a heavy memory restriction that allows one to compute fully developed 3D steady state flows. Second, the TPT decomposition applied for inverse of the Helmholtz and Laplace operators is scalable, while the backward/forward substitutions of sparse and packed *LU* decompositions used in [5] are not scalable.

Performance of the inversed Stokes operator preconditioned Arnoldi iteration (19)-(22) for computation of the leading eigenvalues is illustrated in Table 6. The same benchmark problem as above was considered for two values of the Grashof number $10^7$ and $10^8$. The ARPACK package was implemented [28]. In both cases 16 Krylov basis vectors are sufficient to meet the ARPACK convergence relative convergence criterion of $10^{-6}$. In the first case, $Gr=10^7$, the leading eigenvalue is real and no shift is needed ($\sigma=0$ in Eq. (19)). The numerical experiment shows that iterations converge for the time step $2\leq\delta t\leq40$ and disconverge beyond this interval. For $10\leq\delta t\leq40$ the whole process converges in less than an hour, and again most of CPU time is spent for the inverse of the pressure matrix *C*. The whole process, as well as the convergence of the *C* inverse, significantly slows down when the Grashof number is increased to $Gr=10^8$ and the leading eigenvalue becomes complex. In this case we calculate with the shift $\sigma=(0,0.87)$, which adds another complex time step to each the BiCGstab(2) iteration used for the calculation of Krylov vectors. The convergence of the whole process is observed now for $0.08\leq\delta t\leq1$, which shows that not only optimal time step, but also the convergence yielding one is problem-dependent. The CPU time needed to calculate a single eigenvalue is beyond 10 hours, while the approach of [5] yields the same result in less than one minute. It is stressed again, however, that the present approach removes the memory restrictions and allows for scalable computations, which is not the case of [5].

## 6. Implementation of TPT and TPF in orthogonal curvilinear coordinates

Apparently, the tensor product representation of the Laplacian or Helmholtz operators and factorization (25) cannot always be implemented in general curvilinear orthogonal coordinates. Consider, for example, a Poisson equation in cylindrical coordinates

$$D_{rr}u + \frac{1}{r^2}D_{\theta\theta}u + D_{zz}u = f \tag{41}$$

where $D_{rr} = \frac{1}{r}\frac{\partial}{\partial r}\left(r\frac{\partial}{\partial r}\right)$, $D_{\theta\theta} = \frac{\partial^2}{\partial \theta^2}$, $D_{zz} = \frac{\partial^2}{\partial z^2}$. Assume that the solution domain is covered by a grid $r_i, \theta_j, z_k$, $i = \overline{1,K}, j = \overline{1,M}, k = \overline{1,M}$. In the following, we keep the same notations for the differential and discretized operators. If the eigenvalue decompositions of the discretized operators in $\theta$- and $z$-directions are known

$$D_{\theta\theta} = E_\theta \Lambda_\theta E_\theta^{-1}, \quad D_{zz} = E_z \Lambda_z E_z^{-1} \tag{42}$$

one may look for solution of Eq. (41) in the form

$$u(r_i, \theta_j, z_k) = \sum_{n=1}^{N}\sum_{m=1}^{M} a_{mn}(r_i) E_{\theta,m}(\theta_j) E_{z,n}(z_k) \tag{43}$$

where $E_{\theta,m}$ and $E_{z,n}$ are the $m$-th and $n$-th eigenvectors of $D_{\theta\theta}$ and $D_{zz}$, respectively. Substitution of Eq. (43) into Eq. (41) yields for each coefficient $a_{mn}$

$$\left[D_{rr} + \frac{\lambda_{\theta,m}}{r^2} + \lambda_{z,k}\right] a_{mn}(r_i) = \left(I_r \otimes E_\theta^{-1} \otimes E_z^{-1}\right)f \tag{44}$$

Equations (44) for $a_{mn}$ can be solved, for example, by the Thomas algorithm for banded matrices. Alternatively, we can consider $M$ eigenvalue decompositions of the operators

$$D_{rr} + \frac{\lambda_{\theta,m}}{r^2} = E_{r,m}\Lambda_{r,m}E_{r,m}^{-1}, \quad m = \overline{1,M} \tag{45}$$

Then denoting the diagonal values of $\Lambda_{r,m}$ as $\lambda_{i,m}$, $i = \overline{1,K}$, we obtain

$$u = \sum_{m=1}^{M}\left[(E_{r,m} \otimes E_\theta \otimes E_z)\Lambda^{-1}(E_{r,m}^{-1} \otimes E_\theta^{-1} \otimes E_z^{-1})f\right], \tag{46}$$

and the diagonals values of the matrix $\Lambda$ are $\Lambda_{imk} = \lambda_{i,m} + \lambda_{z,k}$. We can mention here that in the case of general orthogonal curvilinear coordinates the eigenvalue factorization in at least one direction is always possible, so that a 3D problem can be reduced to a series of 2D problems.

## 7. Concluding remarks

In this study we formulated and performed a semi-direct inverse of the Stokes operator using an extended Uzawa method. Our pressure matrix $C$ is an analog of the Uzawa matrix, however not necessarily symmetric or positive semi defined. As in the Uzawa method, the inverse of the pressure matrix $C$ is the main bottleneck of the whole calculation. We have discussed the semi-analytic inverse of the Stokes operator, where only the pressure matrix $C$ is inversed iteratively, as well as fully analytic inverse, where the pressure matrix is treated as "artificially sparse" and a sparse matrix direct solver is applied.

We have shown that factorization of the Stokes operator followed by a fast direct method used to inverse the Helmholtz and Laplacian operators and the preconditioned Krylov subspace iterations used to inverse the pressure matrix, allow one to perform computations with a large time step. The latter is needed for application of inverse Stokes operator preconditioned Newton and Arnoldi methods for calculation of three-dimensional steady states and analysis of their stability. The corresponding test calculations showed that in this way the heavy memory demands of the approach [5] can be removed. On the other hand, the calculations can become fast enough only with a massive parallelization of the basic Laplace/Helmholtz operator direct inverse. The parallelization is algorithmically straight-forward.

We have shown additionally that the tensor product factorization (TPF) method, possibly combined with the Thomas solver (TPT), is sometimes, but not always, faster than iterative methods. Our numerical experiments made for incompressible Boussinesq equations showed that the direct TPF and TPT solvers can perform faster than an iterative method on fine grids and for large Reynolds (Grashof) numbers, when convergence of any iterative method slows down. It is emphasized that since the methods are direct, their computational cost depends only on the problem size, but not on governing parameters, which may make it attractive for cases where iterative methods converge too slowly.

In spite of the fact that the whole TPF/TPT solution is restricted to Poisson and Helmholtz equations defined in the Cartesian coordinates only, the eigenvalue decomposition can be used to reduce the dimension of the problem from 3D to a series of 2D and sometimes, e.g., in cylindrical coordinates, to a series of 1D problems. The effectiveness of the latter is yet to be examined. In the case of, say, vector Laplacian in orthogonal curvilinear coordinates one should take care of cross-conjugate terms when the $x_1$-component of the Laplacian contains $x_2$- and $x_3$-components of the vector along with the $x_1$-component. These terms sometimes can be excluded by a change of variables, as was carried out in [29] for cylindrical coordinates. Alternatively, in time-dependent problems these terms can be treated explicitly.

**Acknowledgement**

We gratefully acknowledge the contribution of Laurette Tuckerman to this work. This research was supported by a grant from the Ministry of Science, Culture & Sport, Israel & the Ministry of Research, France, grant No. 3-4293, and by the German-Israeli Foundation, grant No. I-954 - 34.10/2007.

Table 1. CPU times (sec) consumed by the BiCGstab iterative solver ($t_{BiCG}$), by the tensor product factorization ($t_{TPF}$), and by the tensor product factorization combined with the Thomas algorithm ($t_{TPF}$). Computations for convection of air ($Pr=0.71$) in laterally heated square ($A=1$) and tall ($A=8$) cavities. Calculation on a single scalar Intel 2.4 GHz processor.

| Problem | A=1, Gr=$10^5$, 8700 time steps, $100^2$ grid | | | A=1, Gr=$10^6$, 10,000 time steps, $100^2$ grid | | | A=1, Gr=$10^7$, 10,000 time steps, $100^2$ grid | | | A=8, Gr=$4.8\times10^5$, 17,175 time steps (5 periods), 100×800 grid | | |
|---|---|---|---|---|---|---|---|---|---|---|---|---|
| Variable | $t_{BiCG}$ | $t_{TPF}$ | $t_{TPT}$ | $t_{BiCG}$ | $t_{TPF}$ | $t_{TPT}$ | $t_{BiCG}$ | $t_{TPF}$ | $t_{TPT}$ | $t_{BiCG}$ | $t_{TPF}$ | $t_{TPT}$ |
| T | 17.86 | 51.78 | 29.39 | 21.44 | 60.03 | 33.67 | 21.16 | 60.88 | 33.77 | 535.95 | 3366.8 | 409.5 |
| $v_x$ | 22.77 | 51.22 | 28.61 | 24.33 | 59.50 | 32.95 | 24.31 | 58.72 | 32.47 | 657.10 | 3356.8 | 406.0 |
| $v_y$ | 18.59 | 51.06 | 28.41 | 21.36 | 57.31 | 32.86 | 21.41 | 58.39 | 33.02 | 532.89 | 3305.4 | 399.6 |
| $\delta p$ | 246.83 | 52.31 | 28.67 | 305.82 | 59.83 | 32.80 | 343.5 | 59.55 | 32.98 | 16820 | 3372.13 | 409.8 |
| total | 306.05 | 206.37 | 115.08 | 372.94 | 236.67 | 132.28 | 410.37 | 237.53 | 132.23 | 18546 | 13401 | 1625 |

Table 2. CPU times (sec) consumed for 10,000 time steps by the BiCGstab iterative solver ($t_{BiCG}$) and by the present eigenvalue decomposition solvers ($t_{TPF}$ and $t_{TPT}$) for convection of air (Pr=0.71) in laterally heated cubical cavity. Calculation on a single scalar Intel 2.4 GHz processor.

| Problem | Gr=$10^5$, $30^3$ grid | | | Gr=$10^5$, $50^3$ grid | | | Gr=$10^6$, $50^3$ grid | | | Gr=$10^7$, $50^3$ grid | | |
|---|---|---|---|---|---|---|---|---|---|---|---|---|
| Variable | $t_{BiCG}$ | $t_{TPF}$ | $t_{TPT}$ | $t_{BiCG}$ | $t_{TPF}$ | $t_{TPT}$ | $t_{BiCG}$ | $t_{TPF}$ | $t_{TPT}$ | $t_{BiCG}$ | $t_{TPF}$ | $t_{TPT}$ |
| T | 175.8 | 86.4 | 62.5 | 1004.4 | 611.6 | 428.5 | 920.6 | 608.5 | 426.4 | 929.0 | 611.2 | 428.2 |
| $v_x$ | 183.1 | 85.3 | 61.7 | 1400.1 | 618.6 | 433.4 | 1381.1 | 618.6 | 433.4 | 1393.8 | 619.8 | 434.3 |
| $v_y$ | 193.4 | 84.7 | 61.3 | 1385.6 | 594.8 | 416.8 | 1368.9 | 596.2 | 417.7 | 1379.6 | 598.2 | 419.1 |
| $v_z$ | 125.7 | 84.6 | 61.2 | 881.6 | 600.7 | 420.9 | 872.4 | 599.4 | 420.0 | 879.8 | 604.5 | 423.6 |
| $\delta p$ | 906.4 | 87.1 | 63.0 | 8566.0 | 609.3 | 426.9 | 7393.6 | 609.4 | 427.0 | 7699.5 | 612.3 | 429.0 |
| total | 1584.3 | 428.2 | 309.7 | 13237.7 | 3035.0 | 2126.5 | 11936.7 | 3032.1 | 2124.5 | 12281.8 | 3046.0 | 2134.2 |

| Problem | Gr=$10^5$, $75^3$ grid | | | Gr=$10^6$, $75^3$ grid | | | Gr=$10^7$, $75^3$ grid | | |
|---|---|---|---|---|---|---|---|---|---|
| Variable | $t_{BiCG}$ | $t_{TPF}$ | $t_{TPT}$ | $t_{BiCG}$ | $t_{TPF}$ | $t_{TPT}$ | $t_{BiCG}$ | $t_{TPF}$ | $t_{TPT}$ |
| T | 4042.4 | 3169.4 | 2184.8 | 3260.4 | 3171.6 | 2186.3 | 3264.6 | 3169.3 | 2184.8 |
| $v_x$ | 5409.2 | 3110.1 | 2179.1 | 5402.3 | 3111.6 | 2180.2 | 5398.9 | 3109.3 | 2178.6 |
| $v_y$ | 5394.5 | 3111.0 | 2179.8 | 5394.0 | 3112.9 | 2181.1 | 5373.8 | 3111.59 | 2180.2 |
| $v_z$ | 3114.8 | 3138.8 | 2199.2 | 3122.4 | 3140.9 | 2200.7 | 3124.5 | 3136.51 | 2197.6 |
| $\delta p$ | 38999.1 | 3169.8 | 2221.0 | 34517.4 | 3172.1 | 2222.6 | 34706.7 | 3169.2 | 2220.5 |
| total | 56959.9 | 15699.2 | 10963.9 | 51696.5 | 15709.1 | 10970.9 | 51868.5 | 15695.9 | 10961.7 |

Table 3. CPU times (sec) consumed for 10,000 time steps by the BiCGstab iterative solver ($t_{BiCG}$) and by the present eigenvalue decomposition solvers ($t_{TPF}$ and $t_{TPT}$) for convection of air (Pr=0.71) in laterally heated cubical cavity. Calculation on a single vector Xeon(R) CPU 5355 2.66 GHz processor.

| Problem | Gr=$10^5$, $50^3$ grid | | | Gr=$10^6$, $50^3$ grid | | | Gr=$10^7$, $50^3$ grid | | |
|---|---|---|---|---|---|---|---|---|---|
| Variable | $t_{BiCG}$ | $t_{TPF}$ | $t_{TPT}$ | $t_{BiCG}$ | $t_{TPF}$ | $t_{TPT}$ | $t_{BiCG}$ | $t_{TPF}$ | $t_{TPT}$ |
| T | 76.2 | 29.12 | 20.40 | 76.3 | 28.9 | 20.25 | 76.3 | 29.0 | 20.32 |
| $v_x$ | 132.2 | 27.39 | 19.19 | 133.7 | 27.3 | 19.13 | 135.94 | 27.5 | 19.27 |
| $v_y$ | 131.9 | 27.45 | 19.23 | 133.3 | 27.3 | 19.13 | 135.51 | 27.6 | 19.34 |
| $v_z$ | 72.2 | 27.29 | 19.12 | 42.3 | 27.3 | 19.13 | 72.4 | 27.4 | 19.20 |
| $\delta p$ | 724.7 | 29.32 | 20.54 | 1166.5 | 29.2 | 20.46 | 1222.1 | 29.3 | 20.53 |
| total | 1137.3 | 140.6 | 98.49 | 1582.2 | 140.0 | 98.09 | 1642.3 | 1407 | 98.65 |

| Problem | Gr=$10^5$, $75^3$ grid | | | Gr=$10^6$, $75^3$ grid | | | Gr=$10^7$, $75^3$ grid | | |
|---|---|---|---|---|---|---|---|---|---|
| Variable | $t_{BiCG}$ | $t_{TPF}$ | $t_{TPT}$ | $t_{BiCG}$ | $t_{TPF}$ | $t_{TPT}$ | $t_{BiCG}$ | $t_{TPF}$ | $t_{TPT}$ |
| T | 369.9 | 104.8 | 72.24 | 269.2 | 105.1 | 72.45 | 269.1 | 105.1 | 72.45 |
| $v_x$ | 547.4 | 117.2 | 80.79 | 551.1 | 117.1 | 80.72 | 551.4 | 117.5 | 81.00 |
| $v_y$ | 543.3 | 116.7 | 80.45 | 546.5 | 116.5 | 80.31 | 547.1 | 116.9 | 80.58 |
| $v_z$ | 257.6 | 109.5 | 75.48 | 257.9 | 109.5 | 75.48 | 257.7 | 109.7 | 75.62 |
| $\delta p$ | 3186.4 | 104.4 | 71.97 | 4696.8 | 104.7 | 72.17 | 5454.7 | 104.9 | 72.31 |
| total | 4904.6 | 552.6 | 380.93 | 6321.5 | 552.8 | 381.14 | 7080.0 | 554.0 | 381.97 |

| Problem | Gr=$10^5$, $100^3$ grid | | | Gr=$10^6$, $100^3$ grid | | | Gr=$10^7$, $100^3$ grid | | |
|---|---|---|---|---|---|---|---|---|---|
| Variable | $t_{BiCG}$ | $t_{TPF}$ | $t_{TPT}$ | $t_{BiCG}$ | $t_{TPF}$ | $t_{TPT}$ | $t_{BiCG}$ | $t_{TPF}$ | $t_{TPT}$ |
| T | 872.7 | 384.4 | 262.81 | 877.1 | 383.9 | 262.47 | 6256 | 384.7 | 263.01 |
| $v_x$ | 1298.4 | 354.8 | 242.57 | 1307.4 | 355.1 | 242.78 | 1310.1 | 355.1 | 242.78 |
| $v_y$ | 1293.0 | 357.5 | 244.42 | 1300.3 | 357.7 | 244.55 | 1302.4 | 356.9 | 244.01 |
| $v_z$ | 600.5 | 361.9 | 247.43 | 601.4 | 360.9 | 246.74 | 601.3 | 361.9 | 247.43 |
| $\delta p$ | 9036.4 | 383.7 | 262.33 | 12133.4 | 383.77 | 262.38 | 14152.1 | 384.2 | 262.67 |
| total | 13101.0 | 1842.3 | 1259.55 | 16219.6 | 1841.3 | 1258.92 | 17991.5 | 1842.8 | 1259.89 |

Table 4. CPU times (sec) consumed by the time stepping using inverse of the Stokes operator with iterative or sparse direct inverse of the pressure matrix *C* for the same 2D tests as in Table 1, with 100×100 stretched grid.

| Gr | No. of time steps | $t_{total,bicg\_C}$ | $t_{bicg\_C}$ | $t_{total,mumps}$ | $t_{LU}$ | $t_{bs}$ |
|---|---|---|---|---|---|---|
| $10^5$ | 8,200 | 751 | 611 | 28,618 | 157 | 28,372 |
| $10^6$ | 10,000 | 919 | 755 | 17,137 | 151 | 16,833 |
| $10^7$ | 10,000 | 1021 | 856 | 8,822 | 144 | 8,578 |

Table 5. Number of BiCGstab(2) iterations needed to compute the Newton correction $N_{Newton}$ for the first five Newton iterations and maximal number of BiCGstab(2) iterations $N_C$ needed for the inverse of pressure matrix $C$. Computation of steady state of the benchmark problem of [24],[25] at $Gr=10^7$ starting from the steady state at $Gr=5\times10^6$ as an initial guess. For $\delta t \geq 80$ iterations for the inverse of $C$ disconverge. Single scalar Intel 2.4 GHz processor.

| iteration | 1 | | 2 | | 3 | | 4 | | 5 | | $t_{total}$ |
|---|---|---|---|---|---|---|---|---|---|---|---|
| $\delta t$ | $N_{Newton}$ | $N_C$ | $N_{Newton}$ | $N_C$ | $N_{Newton}$ | $N_C$ | $N_{Newton}$ | $N_C$ | $N_{Newton}$ | $N_C$ | (sec) |
| 1 | 64 | 368 | 21 | 141 | 41 | 229 | 11 | 79 | 1 | 52 | 1312 |
| 2 | 48 | 139 | 31 | 130 | 29 | 72 | 9 | 43 | 1 | 64 | 846 |
| 5 | 61 | 96 | 27 | 70 | 31 | 67 | 9 | 63 | 5 | 55 | 762 |
| 10 | 74 | 96 | 38 | 61 | 17 | 58 | 37 | 50 | 2 | 31 | 793 |
| 20 | 99 | 64 | 34 | 124 | 45 | 55 | 19 | 46 | 10 | 33 | 845 |
| 30 | 94 | 326 | 40 | 300 | 54 | 54 | 35 | 29 | 13 | 24 | 806 |
| 40 | 33 | 122 | 39 | 42 | 38 | 38 | 52 | 28 | 5 | 22 | 720 |
| 50 | 140 | 228 | 40 | 225 | 43 | 41 | 46 | 33 | 18 | 38 | 799 |
| 60 | 132 | 235 | 46 | 51 | 54 | 34 | 51 | 67 | 1 | 19 | 753 |
| 70 | 129 | 135 | 43 | 33 | 45 | 31 | 69 | 112 | 1 | 17 | 729 |

Table 6. Number of Krylov vectors $N_{vectors}$, maximal number of BiCGstab(2) iterations needed to compute the Krylov vectors for Arnoldi iterations $N_{Arnoldi}$ and maximal number of BiCGstab(2) iterations $N_C$ needed for the inverse of pressure matrix $C$. Calculation of the dominant eigenvalues corresponding to the steady state of the benchmark [24],[25]: $\lambda=(-0.02812, 0)$ at $Gr=10^7$ and $\lambda=(-0.032, 0.8663)$ at $Gr=10^8$.

| | $Gr=10^7$ | | | | | $Gr=10^8$ | | | |
|---|---|---|---|---|---|---|---|---|---|
| $\delta t$ | $N_{vectors}$ | $N_{Arnoldi}$ | $N_C$ | $t_{total}$ (sec) | $\delta t$ | $N_{vectors}$ | $N_{Arnoldi}$ | $N_C$ | $t_{total}$ (sec) |
| | | | | | 0.08 | 16 | 834 | 746 | 38013 |
| 2 | 16 | 128 | 931 | 17438 | 0.1 | 16 | 877 | 755 | 38491 |
| 5 | 16 | 99 | 180 | 7603 | 0.2 | 16 | 672 | 903 | 42660 |
| 10 | 16 | 103 | 875 | 6012 | 0.3 | 16 | 592 | 968 | 47491 |
| 20 | 16 | 116 | 801 | 5980 | 0.4 | 16 | 532 | 979 | 50949 |
| 30 | 16 | 141 | 641 | 6091 | 0.5 | 16 | 484 | 945 | 52224 |
| 40 | 16 | 149 | 736 | 5658 | 0.6 | 16 | 480 | 979 | 55974 |